# Broadband mode conversion via gradient index metamaterials

HaiXiao Wang*, [1], YaDong Xu*, [1, †], Patrice Genevet [2, ‡], Jian-Hua Jiang [1, §], HuanYang Chen [1, 3, **]

[1] *College of Physics, Optoelectronics and Energy & Collaborative Innovation Center of Suzhou Nano Science and Technology, Soochow University, No.1 Shizi Street, Suzhou 215006, China*

[2] *Centre de Recherche sur l'Hétéro-Epitaxie et ses Applications, CNRS, Rue Bernard Gregory, Sophia-Antipolis, 06560 Valbonne, France.*

[3] *Key Lab of Advanced Optical Manufacturing Technologies of Jiangsu Province & Key Lab of Modern Optical Technologies of Education Ministry of China, Soochow University, Suzhou 215006, China*

**Abstract:** We propose a design for broadband waveguide mode conversion based on gradient index metamaterials (GIMs). Numerical simulations demonstrate that the zeroth order of transverse magnetic mode or the first order of transverse electric mode ($TM_0/TE_1$) can be converted into the first order of transverse magnetic mode or the second order of transverse electric mode ($TM_1/TE_2$) for a broadband of frequencies. As an application, an asymmetric propagation is achieved by integrating zero index metamaterials inside the GIM waveguide.

**Introduction**

Along with the rapid development of photonic integrated circuits, the mode control techniques including mode filtering, mode separation, and mode conversion become crucial for designs of integrated optical systems. Among these techniques, mode conversion plays a more fundamental role and thus received a considerable research interest [1-9]. There have been several methods to achieve mode conversion. Usually, an excellent mode converter is characterized by large bandwidth, low loss and extinction ratio. For example, it has been suggested that an ultra-compact interferometer formed by nano-waveguides can function as mode converter by introducing an optical phase difference [3]. Such an ultracompact structure features low loss, but with polarization-dependence and narrow bandwidth as its shortage. In Ref. [9], a similar geometry has been proposed for broadband mode conversion based on interference effect of light propagating through two differential dielectric waveguides. In both systems, the key is to recombine light after introducing a phase difference between two different channels. However, beam splitters at the beginning and at the end of such devices generally induce significant backscattering that reduces the efficiency of mode conversion. Suppressing the backscattering is therefore essential to improve the performance of interferential mode converters.

Nanostructured optical interfaces -or metasurfaces- have recently opened new avenues for manipulating light properties at interfaces [10-11]. In particular, metasurfaces based on gradient index metamaterials (GIMs) have been proposed to completely convert propagating waves into surface-like waves [12]. Waveguides with symmetric GIMs can convert the propagating mode (PM) gradually into surface-like mode (SM) with negligible scattering, achieving asymmetric propagation and waveguide cloaking for a broadband of frequencies independent of the polarization of the incident wave [13-14]. It is therefore desirable to utilize GIMs to design a waveguide structure for mode conversion with suppressed backscattering and high mode conversion efficiency. In this work, we will introduce asymmetric GIMs into the waveguide, and theoretically demonstrate that mode conversion between two lowest waveguide modes can be achieved for a broad bandwidth of frequencies and independent of incident polarizations.

* These authors contribute equally to this work.
† ydxu@suda.edu.cn
‡ patrice.genevet@crhea.cnrs.fr
§ joejhjiang@sina.com
** chy@suda.edu.cn

**Results and Discussions**

The schematic diagram of the waveguide with asymmetric GIMs is shown in Fig. 1a. It is a 1D parallel-plate waveguide with two GIMs attached to its outer perfect electric conductor (PEC) walls. The index profiles of the two GIMs along $x$ direction are different from each other and can be described as

$$\varepsilon_i(x) = \mu_i(x) = \begin{cases} 1 + g_i \dfrac{\kappa(x+L/2)}{2k_0 d}, & -\dfrac{L}{2} \leq x \leq 0, i = 1, 2 \\ 1 - g_i \dfrac{\kappa(x-L/2)}{2k_0 d}, & 0 \leq x \leq \dfrac{L}{2}, i = 1, 2 \end{cases}, \quad (1)$$

where $\kappa=0.2\,k_0$ ($k_0$ is the wave vector in free space) is a predesigned momentum parameter, $d=1.5$mm is the thickness of the GIMs along $y$ direction. The length of the waveguide is $L=240$mm, $w=22.5$mm is the distance between the two parallel GIMs. Furthermore, $g_1$ and $g_2$ are referred as the gradient factors of the upper and lower GIMs, respectively. We will show that the mode conversion effect can be tuned by adjusting the values of $g_1$ and $g_2$ as well as the working frequencies. The difference between $g_1$ and $g_2$ is crucial for mode conversion. A classical geometric optics interpretation will be introduced later. The mode conversion effect is actually due to the difference of the accumulated phases during the light propagation along the upper and lower GIMs. To depict the differences between the two GIMs, their index profiles are plotted in Fig. 1b. As long as the refractive index does not vary too quickly, the waveguide structure induces negligible backscattering. For this reason we set both gradient factors around unity. In view of this configuration, the index profiles of GIMs in Ref. [14] is a special case with $g_1=g_2=1$.

To visualize the mode conversion effect, we perform numerical simulations by using the COMSOL Multiphysics. Both TM and TE polarizations are investigated. For TM polarization, the $TM_0$ mode with a frequency of 9.5GHz is incident from the left port. When both upper and lower GIMs have the same gradient factor ($g_1= g_2$, i.e. they are symmetric, which has been used for waveguide cloaking in Ref. [14]), the output waveguide mode keeps the same as the input waveguide mode ($TM_0$), as shown in Fig. 1c. However, when the gradient factor of the upper GIM $g_1$ is slightly modified and the other gradient factor $g_2$ is kept unchanged, mode conversion from $TM_0$ to $TM_1$ can be achieved. Fig. 1d shows that, when $g_1$ and $g_2$ are set as 1.015 and 1, respectively, a nearly complete waveguide mode conversion is realized at a frequency of 9.5GHz: the output mode turned into $TM_1$ mode completely. For the case of TE polarization, similar results can be found. For example, in Fig. 1e, a $TE_1$ mode with a frequency of 12GHz is incident from left, when the upper GIM slab and lower GIM slab have the same gradient factor, the same $TE_1$ mode will come out from the output port. However, when the gradient factor of the upper GIM slab $g_1$ is tuned to 1.018 while the gradient factor of the lower GIM slab $g_2$ remains to be unity, the output mode will be converted into a $TE_2$ mode. It should be noted that due to the low gradient factors, there is little backscattering for the waveguide, unless certain Fano resonances of higher modes are excited [15], which we will discuss in the Appendix. These results clearly demonstrate that our device operates as a mode converter for both TM and TE polarizations.

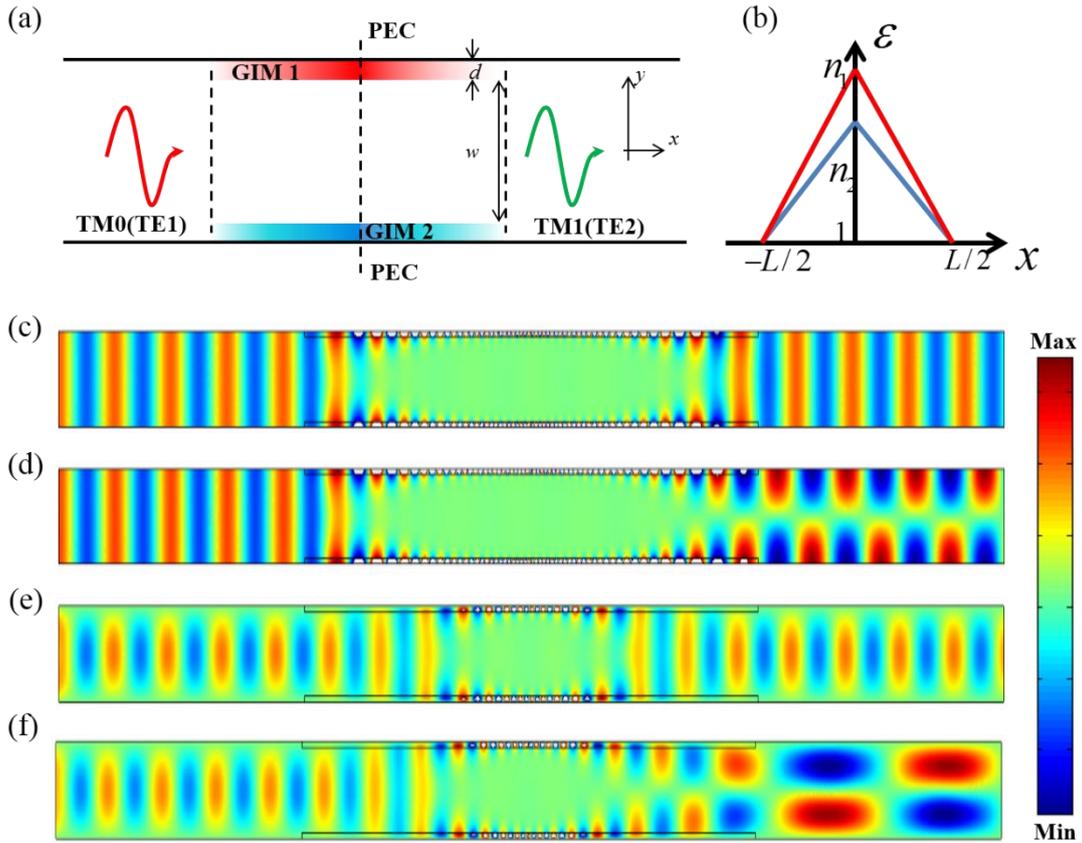

*Figure 1* *(a) The schematic plot of the waveguide mode converter. (b)The gradient index profile of the two GIM slabs, where $n_1$ is the maximum refractive index of the upper GIM slab, while $n_2$ is the maximum refractive index of the lower GIM slab. (c, d) The simulated magnetic field patterns for $TM_0$ mode incident from left to right at 9.5GHz with (c) $g_i=1$ (i=1, 2) and (d) $g_i$ (i=1,2) equal to 1.015 and 1, respectively. The output mode in (d) is converted into a $TM_1$ mode. (e, f) The simulated electric field patterns for $TE_1$ mode incident from left to right at 12GHz with (e) $g_i=1$ (i=1,2) and (f) $g_i$ (i=1,2) equal to 1.018 and 1, respectively. The output mode in (f) is converted into a $TE_2$ mode.*

In the frequency region considered in this work, only two lowest modes are supported in the output port of the waveguide. Therefore the output electromagnetic wave generally consists of two parts. For TM polarization, these are the magnetic fields $H_0$ and $H_1$ for the zeroth order and the first order of eigenmodes, respectively. That is,

$$H = \alpha H_0 + \alpha' H_1 \qquad (2)$$

For TE polarization, they are the electric fields $E_1$ and $E_2$ for the first order and the second order of eigenmodes, i.e.,

$$E = \gamma E_1 + \gamma' E_2 \qquad (3)$$

The coefficients $\alpha$, $\alpha'$, $\gamma$, and $\gamma'$ describe the amplitudes of these eigenmodes at the output port. Thus $|\alpha|^2$ and $|\gamma|^2$ refer to the fraction of output $TM_0$ or $TE_1$ mode, which is closely related to the conversion efficiency. Generally, the conversion efficiency is defined as $\eta=T\cdot f$, where $T$ is the transmission of both modes, and $f$ is the fraction of $TM_1$ or $TE_2$ mode, i.e., $f=|\alpha'|^2/(|\alpha|^2+|\alpha'|^2)$ or $f=|\gamma'|^2/(|\gamma|^2+|\gamma'|^2)$. Thanks to the small gradient factors, the PM can be gradually transferred into SM with suppressed backscattering, indicating that the transmission in our waveguide system is mostly close to unity.

In the following, we show how to continuously tune output mode as function of the frequency by adjusting both gradient factors $g_1$ and $g_2$. In order to explore the parameter space through which mode conversion efficiently happen, the values of $|\alpha|^2$ and $|\gamma|^2$ are numerically calculated for different variables, such as the gradient factors of GIMs and the working frequencies. Firstly, let us explore the case of TM polarization. Figure 2a shows the varying fraction of the output $TM_0$ mode along with the varying gradient factors of the two GIM slabs, where the working frequency is fixed at 6GHz. It is consistent with intuition that either changing $g_1$ or $g_2$ have the same effects, as clearly shown in Fig. 2a. The diagram is symmetric along the diagonal line. Two dash lines in the blue regions refer to the contour line of $|\alpha|^2$ equal to zero, i.e., there is no $TM_0$ mode in the output port (all of them have been converted to $TM_1$ mode). Moreover, we observed that the mode converter works over a broad range of frequencies. Figure 2b shows the relationship between the fraction of the output $TM_0$ mode and the gradient factor of the lower GIM slab ($g_2$) as well as the working frequencies by fixing $g_1=1$. The diagram appears to be periodic along with the varying gradient factor of the lower GIM slab. Point A refers to the case mentioned in Fig. 1d, which indicated that the output mode convert into $TM_1$ mode completely. As the frequency increases, the fraction of the output $TM_0$ mode tends to change more slowly. In particular, when $g_2$ is fixed at 1.015 or 0.985, and the working frequencies range from 9 GHz to 11.7GHz, the fraction of output mode $|\alpha|^2$ tend to be a nearly-constant and close to zero, as revealed by the two white dash lines in Fig. 2b. Hence, the working frequencies have a 26% bandwidth (2.7GHz). For details, we plot the transmission curve for $g_1=1.015$, and $g_2=1$ in Fig. 2c for a finer frequency resolution, where we find that indeed high transmission happens for a broad band of frequencies. One may extend the converting frequency to a higher value, however, it should be noted that higher order modes will be excited for both ports, which will influence the conversion efficiency and more physical scenarios should be taken into consideration.

As for TE polarization, similar effects are observed but compared with $TM_0$ mode, $TE_1$ mode has higher working frequencies in the same waveguide structure. Therefore, we limit the gradient factors range from 0.97 to 1.03 and keep the working frequency range from 12GHz to 16.8 GHz, ensuring that the output port can only support $TE_2$ mode and no higher order modes. Figure 2d present the relationships between the fraction of output $TE_1$ mode and various gradient factors of the two GIM slabs, which also appears to be periodic and symmetric. Point B refers to the case mentioned in Fig. 1f, which indicated that the output mode convert into $TE_2$ mode completely. In order to explore whether such a mode converter works for a broadband of frequencies, we plot the fraction of output $TE_1$ mode with varying working frequencies and the gradient factor of the lower GIM slab ($g_2$) by fixing $g_1=1$, as shown in Fig. 2e. For $g_2 = 0.99$, we find that the fraction of output mode $|\gamma|^2$ tends to be a nearly-constant and close to zero, as revealed by the white dash line. The working frequencies now range from 14GHz to 16.8GHz, showing an 18% bandwidth (2.8GHz). We then plot the transmission from 14GHz to 16.8GHz for the case of $g_1=1$ and $g_2=0.99$ in Fig. 2f for a finer frequency resolution. Except for several weak dips, the transmission is high for a broadband of frequencies. Even for the frequencies of dips, the transmission is above 0.7; therefore the effect will not be compromised too much.

There are several special areas denoted by white dash circles in the diagram, which shows that there are some strong transmission dips. Such dips come from Fano resonances, which we will discuss in details in Appendix. In spite of these special areas, broadband mode converter via GIMs without polarization limitation is still clearly verified.

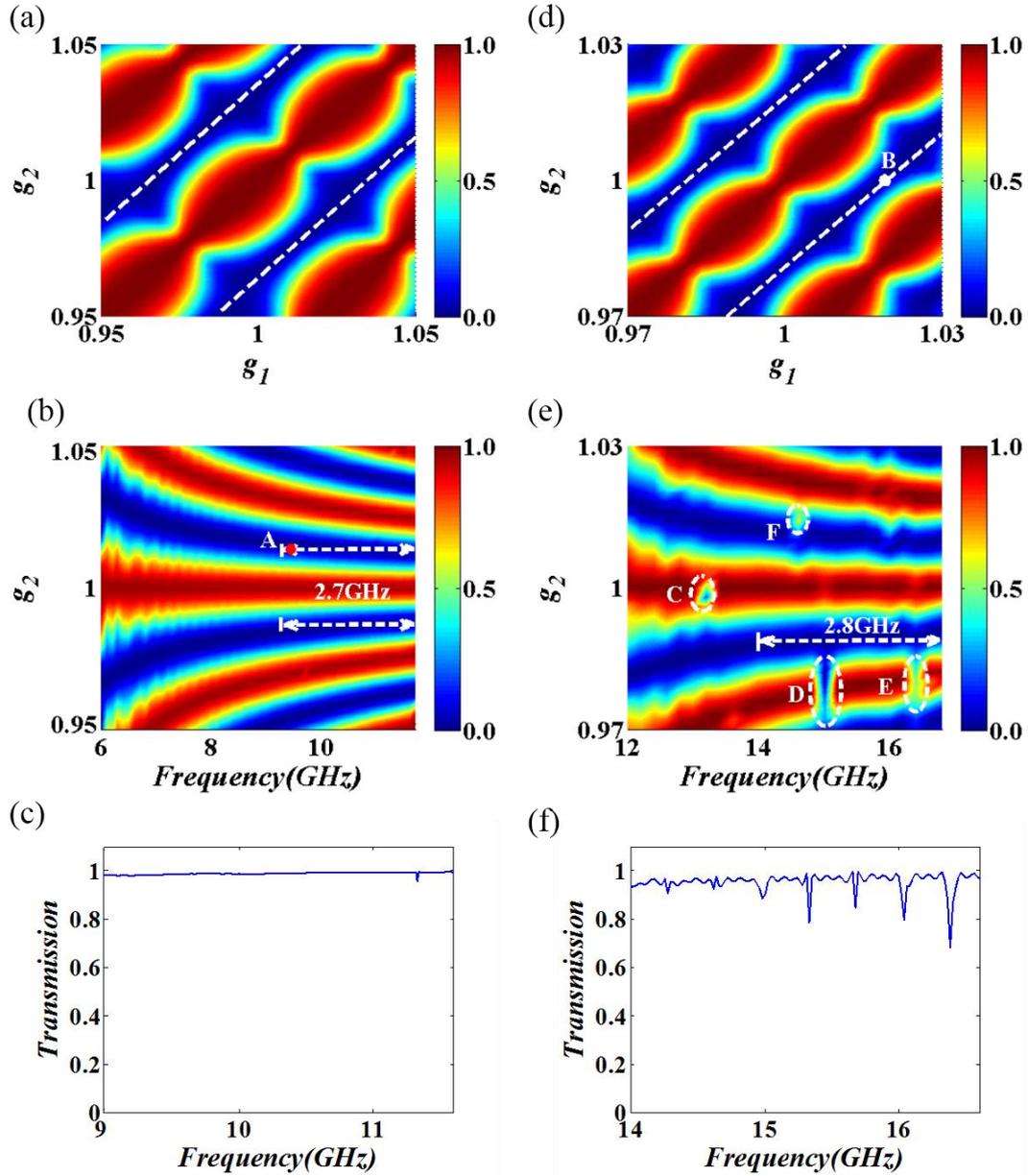

*Figure 2* The characteristics of mode converter with different variables. (a, b, c) For TM polarization, (a) the changing tendency of $|\alpha|^2$ with different gradient factors of GIM slabs for a fixed working frequency at 6GHz. (b) The changing tendency of $|\alpha|^2$ with different working frequencies as well as the gradient factor of the lower GIM slab. The gradient factor of the upper GIM $g_1$ is fixed at 1. Point A in (b) corresponding to the case mentioned in Fig. 1d. Two white dash lines in (b) refer to the broadband mode converting regions with the fluctuation limited to 5%, where $g_2$ are set as 1.015 and 0.985, respectively. The bandwidth is about 2.7GHz. (c) Transmission of the case with $g_1=1.015$, $g_2=1$, (d, e, f) For TE polarization, (d) the changing tendency of $|\gamma|^2$ with different gradient factors of GIM slabs for a fixed working frequency at 12GHz. Point B in (d) corresponding to the case mentioned in Fig.1f. (e) The changing tendency of $|\gamma|^2$ with different working frequencies as well as the gradient factor of the lower GIM slab. The gradient factor of the upper GIM $g_1$ in both cases is fixed at 1. One white dash line in (e) shows the broadband mode converting functionality for a broad bandwidth of about 2.8 GHz. Four special areas denoted by dash white circles (C, D, E, and

*F) in (e) indicated the transmission dips caused by Fano resonances. (f) Transmission of the case with g1=1, g₂=0.99.*

Now, let's examine the mechanism behind the proposed mode converter. As mentioned in Refs. [13, 14], the GIM slabs can convert a PM to a SM with a nearly 100% efficiency, causing little scattering. This can be achieved because the band branch of $TM_0/TE_1$ mode goes below the light line as the refractive indexes of dielectrics of GIM slabs increase (see Fig. 2e and 2f of Ref. [13]). Here for the mode converter, similar physics happens. The only difference is that owing to the asymmetric gradient factors, there is a phase difference of surface modes in the upper and lower GIM slabs. For simplicity, the wave vector of SM ($\beta$) at a specific position of each GIM slab is approximately equal to the wave vector in a bulk media with a dielectric constant equal to that of each GIM slab at the position, which can be viewed as $\beta=2\pi f \cdot n(x)/c$. The changing of refractive index $n$ gives rise to different wave vectors and consequently, different accumulated phases during light propagation in the upper and lower GIM slabs respectively. The accumulated phase difference between these two GIM slabs can be defined as $\theta = \int_{-L/2}^{L/2} \Delta\beta dx$. If $\theta$ equal to $2n\pi$ *(n=0, 1, 2…)*, there is no phase accumulations between upper and lower GIM slabs, i.e. the output mode keep the same as the input mode ($TM_0$ or $TE_1$ mode). While if $\theta$ equal to $(2n+1)\pi$ (n=0, 1, 2…), the optical field through the upper GIM has opposite sign relative to the optical field through the lower GIM. At this condition the output mode turn to the higher mode ($TM_1$ or $TE_2$ mode). Figure 3 represents the fraction of output mode ($TM_1$ and $TE_2$) and the phase difference $\theta$ as a function of $g_2$ with the fixed $g_1$ and fixed working frequency, or as a function of working frequency with the fixed $g_1$ and $g_2$.

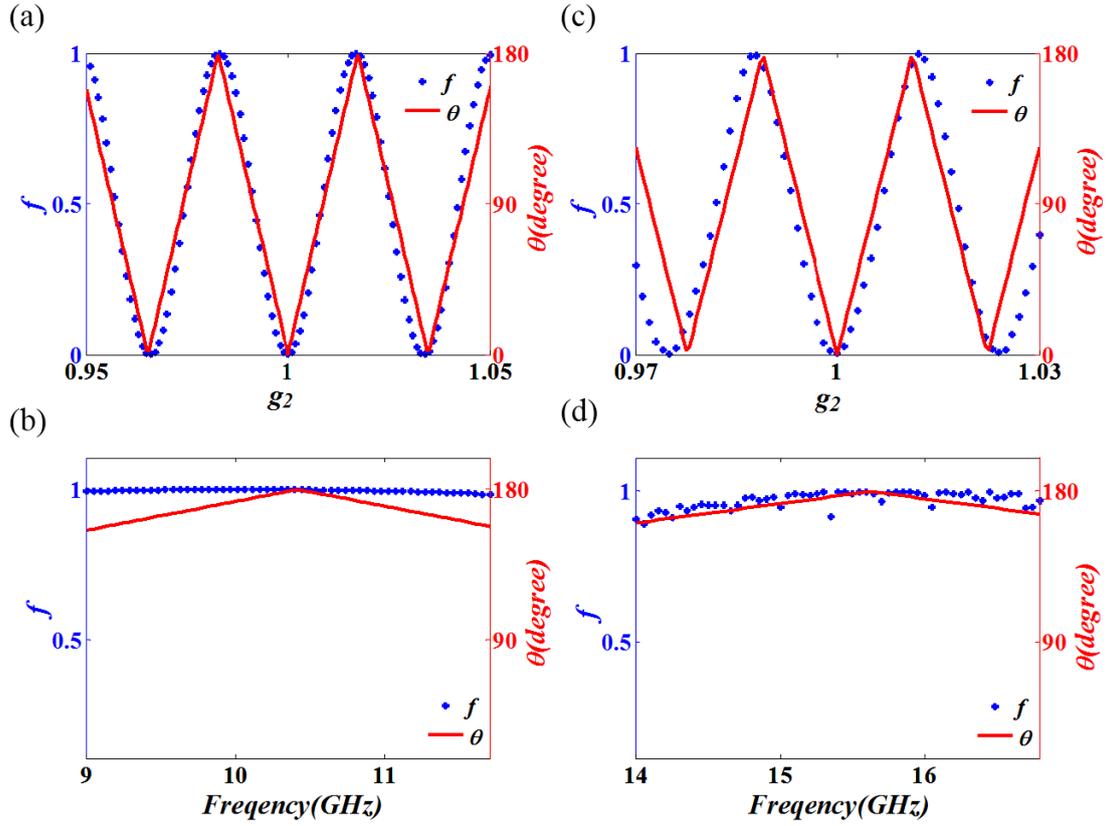

**Figure 3** *The illustration of accumulated phase difference and the fraction of output mode with varying gradient factors and working frequencies. (a, b) For TM polarization, red solid line refers to the accumulated phase difference between two GIM slabs, blue star data points refer to the fraction of $TM_1$ mode, (a) with $g_1$ and the working frequency fixed at 1 and 6GHz respectively, (b) with $g_1$ and $g_2$ fixed at 1.015 and 1, i.e., the set-up of broadband converter for TM polarization. (c, d) For TE polarization, red solid lines refer to the accumulated phase difference between two GIM slabs, blue star data points refer to the fraction of $TE_2$ mode, with $g_1$ and working frequency fixed at 1 and 12GHz, (d) with $g_1$ and $g_2$ fixed at 1 and 0.99, i.e., the set-up of broadband converter for TE polarization. Both (a,c) and (b,d) demonstrate that mode conversion is essentially controlled by the accumulated phase difference between the two asymmetric GIM slabs.*

In order to show the periodically changing fraction of output mode and the phase difference in the same figure, we limit the phase difference from 0 to 180 degree. Note that both TE and TM mode conversion can be explained with the same physical mechanisms, we are discussing TM polarization only and list the result of TE polarization without detailed explanations. As show in Fig. 3a, when there is no accumulated phase difference (i.e. $\theta$ equal to 0 degree), the fraction of the output $TM_1$ mode keeps its minimum value thus no mode conversion occurs, while if the accumulated phase difference reach to its peak (i.e. $\theta$ equal to odd integer multiply of $\pi$), the fraction of output $TM_1$ mode increased to unity, meaning that mode conversion is maximized. It should be emphasized that this explanation is applicable to the case that both gradient factors are around unity. Otherwise, the explanation may not function very well, as also revealed in Fig. 3a. When $g_2$ equal to 0.95 or 1.05, the result is not so exact as that when $g_2$ equal to unity. Regardless of it, both two parameters agree with each other very well with the varying gradient factor. It verifies our explanation that the changing of output mode is stemmed from the asymmetric GIM slabs, which can approximately explained by the accumulated phase difference between these two GIM slabs. The same phenomenon is found for TE polarization (see in Fig. 3c). We then fixed $g_1$ and $g_2$, and plot the accumulated phase difference and the fraction of the output mode at different working frequencies in Fig. 3b (TM) and 3d (TE), for the broadband mode converter mentioned in Fig. 2c and 2f. We find that the fraction of the output mode coincides with the accumulated phase difference very well for a broadband of frequencies for both polarizations. This also explain why the mode converter can work in a broadband of frequencies. As both gradient factors are very close to unity, the accumulated phase differences will not deviate too much from $\pi$ for the working frequencies. However, we should emphasize that the accumulated phase difference method is only approximately correct. When the distance between the two GIMs becomes smaller, the coupling effect will be stronger and the mode conversion will be compromised.

As an application, we combine the properties of our mode converter with those of a slab of zero index metamaterials (ZIMs) to achieve asymmetric propagation in a waveguide. ZIMs have many interesting properties. For example, they can be used to enhance the directive emission for an embedded source [16], or to squeeze electromagnetic waves in a narrow waveguide with ZIMs [17-18]. In addition, total transmission and total reflection can be achieved by introducing defects in ZIMs [19-21]. Figures 4a and 4b schematically show the transverse momentum conservation for reflection and transmission at the interface air/ZIMs. The bigger circle represents the isofrequency curve in air while the smaller one denotes that in ZIMs. The red arrow in the bigger circle is at the critical incident angle, which is nearly horizontal. For the incident angle larger than the critical angle, the ray cannot transmit into ZIMs and has a total reflection at the interface. To be more precise, when the $TM_0$ mode is incident on a ZIM surface, it will propagate through ZIMs as it was a transparent media with a zero reflection coefficient. While $TM_1$ mode are totally reflected, as shown in Fig. 4b. Therefore, putting a ZIM block into the right port of our waveguide structure together with the asymmetric GIM slabs, we achieve asymmetric propagation for TM polarization. Simulations are performed to demonstrate the asymmetric propagation, the related parameters are set as follow: $g_1$=1, $g_2$=1.015, and working frequency is 9GHz. As shown in Fig. 4c, when the $TM_0$ mode with a frequency of 9GHz is incident from the right port, it will first pass through the ZIM and then continue to propagate in the GIM waveguide, leaving the left port as a $TM_1$ mode. However, when the $TM_0$ is incident from the left port, it will first be converted into a $TM_1$ mode after passing through the GIM part. Then it experiences total reflection after reaching the ZIM part, demonstrating asymmetric propagation as shown in Fig. 4d.

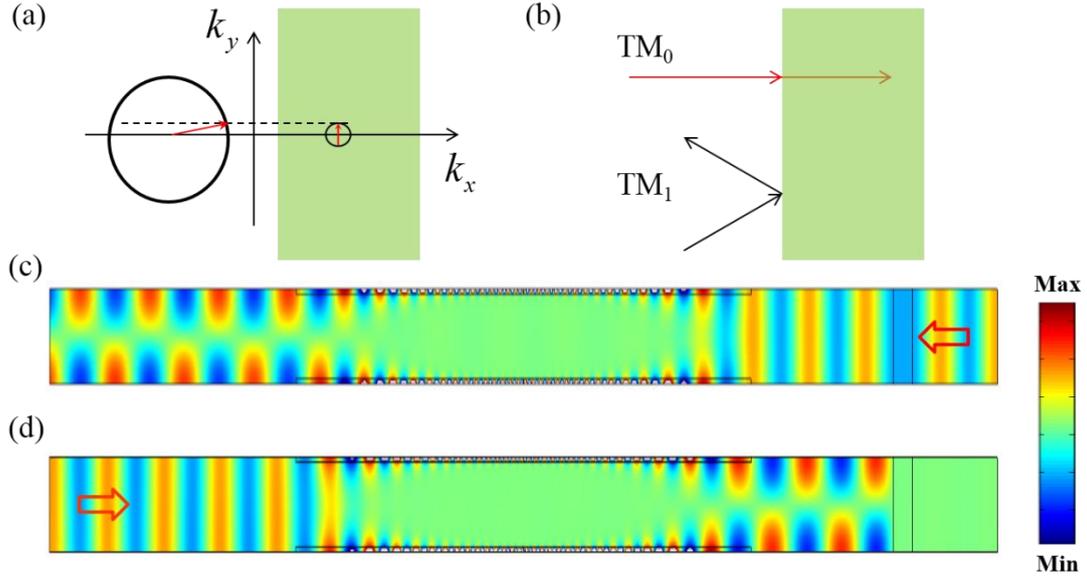

*Figure 4 Asymmetric propagation based on waveguide mode conversion coupled with ZIMs. (a, b) The schematic diagram of reflection and refraction at the interface of air and ZIMs (indicated by green areas). (c, d) The simulated magnetic field patterns for $TM_0$ wave in the waveguide structure with ZIMs. (c) Refers to the case where the wave is incident from right to left, while (d) refers from left to right. Other parameters are set as follows: $g_1=1.015$, $g_2=1$, and $f=9GHz$.*

**Conclusions**

In this paper, we have proposed a broadband mode converter which consists of a waveguide with gradient index metamaterials coatings on its sides. We have studied numerically and analytically the efficiency of mode conversion from $TM_0/TE_1$ into $TM_1/TE_2$ as function of metamaterial gradient factors and working frequencies. In particular, we have shown that our device can maintain relatively high mode conversion over a broad range of frequencies due to the slowly varying refractive index profiles of gradient index metamaterials. We have explained, using simple arguments, that the accumulated phase difference caused by light propagation in the asymmetric GIMs controls waveguide mode conversion. In addition, we have proposed an interesting application of asymmetric TM propagation by introducing zero index metamaterials at one port of the GIM waveguide.

**Acknowledgements**

This work is supported by the National Science Foundation of China for Excellent Young Scientists (grant no. 61322504), the Foundation for the Author of National Excellent Doctoral Dissertation of China (grant no. 201217), and the Priority Academic Program Development (PAPD) of Jiangsu Higher Education Institutions. P. G. is supported by the European Research Council (ERC) under the European Union's Horizon 2020 research and innovation programme (grant agreement FLATLIGHT No 639109). J. H. J. thanks support from the start-up funding of Soochow University.

**Appendix**

For some cases of TE polarization, the system exhibits strong Fano resonances. For example, for "C", "D", "E" and "F" in Fig. 2e, the transmission reaches to values near zero. We plot the transmission for the case of "C", "D", "E", and "F" in Fig. 5a, 5b, 5c, and 5d, respectively. We find that such cases show clear asymmetric resonant line shapes, i.e. they come from Fano resonances.

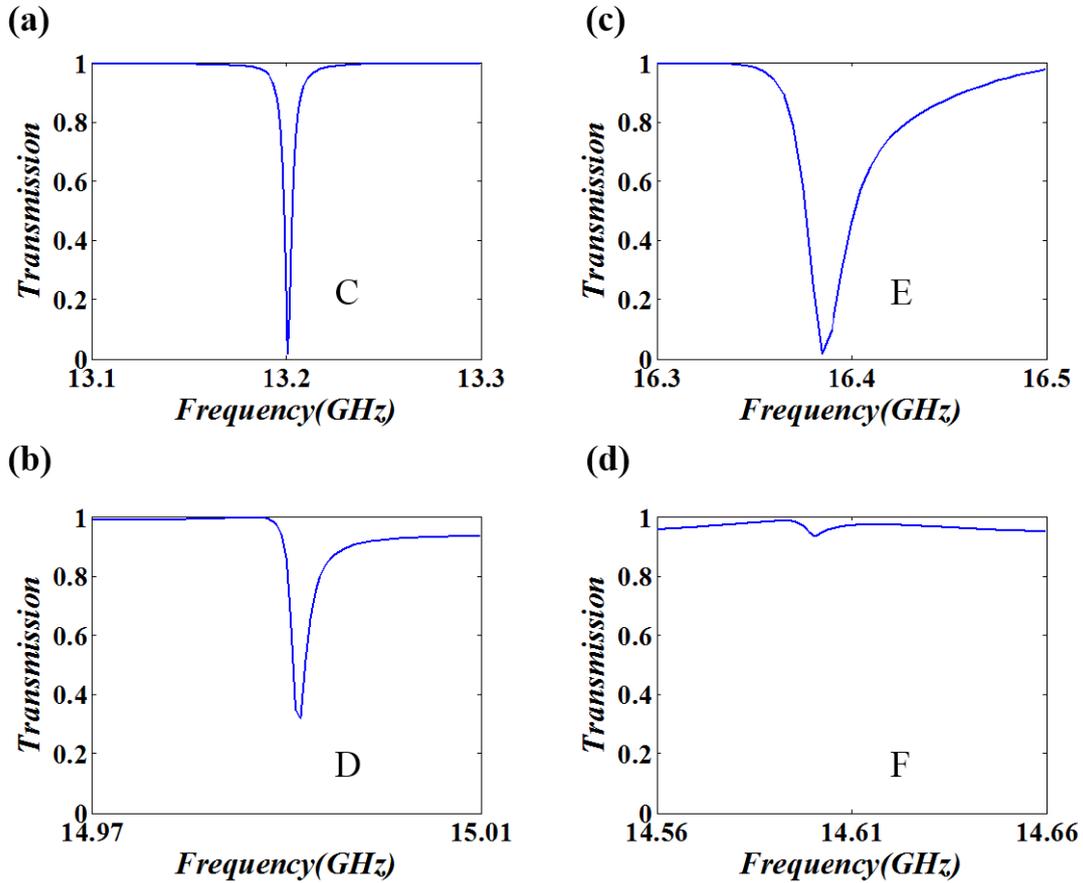

*Figure 5*. *Transmission of the waveguide with different gradient factors for TE polarization, (a) for the case of "C" in Fig. 2e, (b) for the case of "D" in Fig. 2e, (c) for the case of "E" in Fig. 2e, (d) for the case of "F" in Fig. 2e.*

Let us first examine the field patterns at the above transmission dips in Fig. 6, where we find that all of these field patterns share the same characteristic: higher order modes are excited at the middle part of the waveguide and large amount of energy is confined in the area of waveguide with GIM slabs.

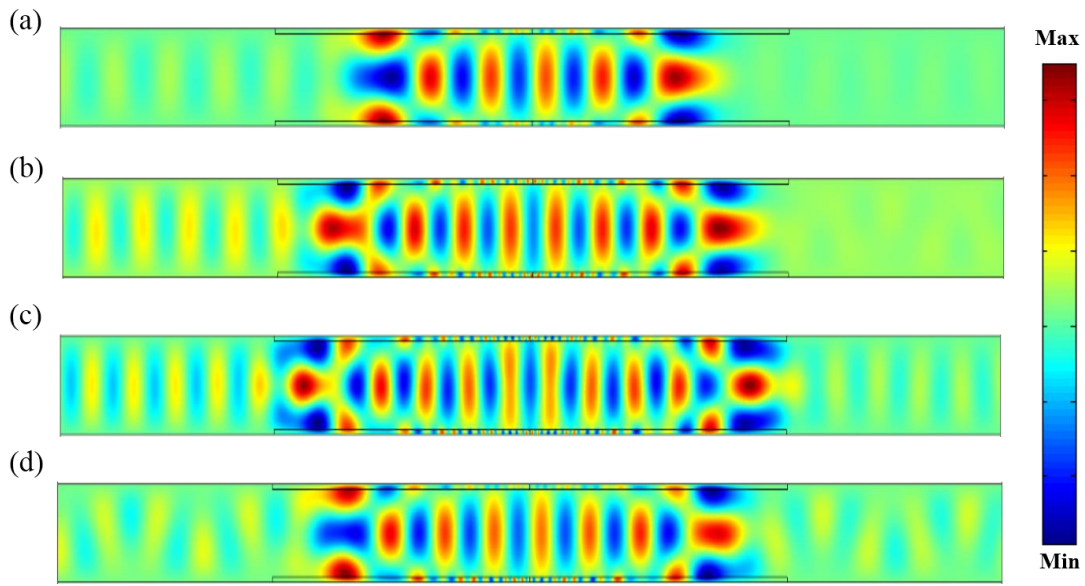

*Figure 6*. Electric field patterns of different cases, (a) refer to case "C" at the resonant frequency of 13.2GHz, (b) refer to case "D" at resonant frequency of 15GHz, (c) refer to case "E" at resonant frequency of 16.4GHz, (d) refer to case "F" at resonant frequency of 14.6GHz.

To explore further, we plot the dispersions of case "C" mentioned above in Fig. 6 (with symmetric GIMs) to explain the underlying physics.

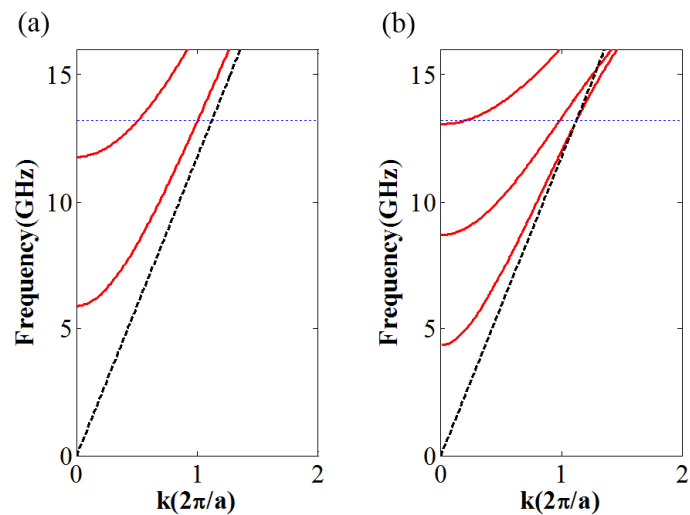

*Figure 7* The dispersion relations for TE mode with the GIMs replaced by different dielectrics. (a) The dispersion relation for TE mode by setting the refractive index of the dielectrics as 1, i.e. an empty waveguide. (b) The dispersion relation for TE mode by setting the refractive index of the dielectrics as 4. Both red solid curves refer to different modes

*supported in the waveguide, black dash lines are the light lines, and dash blue lines refer to the working frequencies of 13.2GHz.*

As shown in Fig. 7, when the TE wave with a frequency of 13.2GHz incident into the waveguide, both $TE_1$ and $TE_2$ modes can be supported. However, due to the symmetry of the system, the $TE_2$ mode (asymmetric mode) could not be excited and there is only $TE_1$ mode in the waveguide. When this $TE_1$ mode continues to propagate through the area of the waveguide with higher dielectrics, the dispersion is totally different, as shown in Fig. 7b. The waveguide can now support higher symmetric mode ($TE_3$ mode), which has already been found in the field patterns in Fig. 6. Thus, two symmetric modes ($TE_1$ and $TE_3$ modes) coexist in the area of the waveguide with dielectrics. However, we should keep in mind that $TE_3$ does not always exist in the whole waveguide structure. As the refractive index of the dielectrics turns to be lower, $TE_3$ mode cannot transmit through the output ports of waveguide. In fact, $TE_3$ mode will experience an open cavity and emit energy in forms of $TE_1$ mode at resonances. This part of energy will interfere with original $TE_1$ mode that already propagate in the waveguide system, forming Fano resonances, see more details in Ref. [15].

Hence the above sharp dips for the special areas denoted by white dash circles in Fig. 2e come from Fano resonances. In fact, the weak dips in Fig. 2f also share the same physics, and the resonance becomes stronger and stronger, as the frequency increases.